\documentclass[a4paper]{jpconf}
\usepackage{graphicx}
\begin{document}
\title{Size dependence of thermoelectric power of Au, Pd, Pt nanoclusters  deposited onto HOPG surface}

\author{P V Borisyuk$^1$, V I Troyan$^1$, Yu Yu Lebedinskii$^{1,2}$, O S Vasilyev$^1$}

\address{$^1$ National Research Nuclear University MEPhI (Moscow Engineering Physics Institute), 31~Kashirskoe sh., Moscow, 115409, Russian Federation}
\address{$^2$ Moscow Institute of Physics and Technology, 9 Institutskiy per., Dolgoprudny, Moscow Region, 141700, Russian Federation}
\ead{osvasilyev@mephi.ru}

\begin{abstract}
The paper presents the study of tunnel current-voltage characteristics of Au, Pd and Pt nanoclusters deposited onto the highly oriented pyrolytic graphite (HOPG) surface by pulsed laser deposition. The analysis of tunnel current-voltage characteristics  obtained by scanning tunneling spectroscopy (STS) allowed to recover the thermoelectric power value of nanoclusters. It was found that the value of thermoelectric power of pulsed laser deposited nanoclusters depends on nanocluster material and shows qualitative difference in size dependence for different materials of nanoclusters. Thus, thermoelectric power value of PLD Pd nanoclusters decreases with increasing of nanocluster size, while for Au nanoclusters this value increases. The analysis of the results are discussed.
\end{abstract}

\section{Introduction}
The temperature gradient in solids causes the formation of an electrical potential difference. The effect performs direct transformation of thermal energy to electrical one. However, the development of new generation of thermoelectric generators based on quantum effects is complicated \cite{Sootsman2009,Zebarjadi2012}. 
The main difficulty while caring out an experiments on measuring the thermoelectric properties of systems of nano-sized scale is the invasiveness of measurements. That means that the applied temperature gradient causes local changes near the contact and should be taken into account while calculating the final result. Furthermore, the electrical potential difference caused by the temperature gradient is very sensitive to the geometry of the contact. Some attempts aimed to solve the problems resulted in sufficient  improvement of thermoelectric figure of merit of bulk materials though were not good enough for practical purposes (see Refs. \cite{Seol2007,Date2014}). In this way the study of thermoelectric properties of individual nanoobjects such as nanoclusters, nanotubes, nanofilms etc.  is rather urgent.

In this paper for thermoelectric power value determination of Pd and Pt nanoclusters we used the original non-contact technique of recovery of thermoelectric power value of metal nanoclusters deposited on a surface of a conductive substrate  \cite{Borisyuk2014,Troyan2014}. The technique is based on the analysis of tunnel current-voltage characteristics ($IV$-curves) obtained by scanning tunneling spectroscopy (STS). An advantage of this technique is the lack of necessity of temperature measurements and samples heating, as well as the possibility of local study of single nanoclusters formed on the surface of the conductive substrates.

Thermoelectric power value of Pd and Pt nanoclusters formed by a pulsed laser deposition technique (PLD) on highly oriented pyrolytic graphite (HOPG) surface (0001) at room temperature was recovered and comparison with previous results on thermoelectric power value of Au nanoclusters \cite{Borisyuk2014,Troyan2014} was made. It was found that the value of thermoelectric power of pulsed laser deposited nanoclusters depends on nanocluster's material and shows qualitative difference in size dependence for different materials of nanoclusters. 
Thermoelectric power value of Au and Pd nanoclusters of volume more than 10~nm$^3$ is typical for bulk materials, whereas for nanoclusters of volume of about  0.1~nm$^3$ the absolute values of thermoelectric power of Au and Pd are several times greater in the absolute value compared with bulk materials.
Thermoelectric power value of PLD Pd nanoclusters decreases with increasing of nanocluster's sizes, while for Au nanoclusters this value increases. Thermoelectric power value of Pt nanoclusters of studied volumes is close to bulk value. 

\section{Material and methods}
Samples with Pd and Pt nanoclusters deposited onto HOPG surface were prepared at the high vacuum conditions ($p\simeq 5\cdot 10^{-10}$ Torr) at room temperature with the use of PLD technique implemented on the basis of XSAM-800 (Kratos) ultra high vacuum (UHV) spectrometer (the detailed description of the experimental technique can be found in Ref.~\cite{Borman2006Eng}) For this purpose the radiation from a YAG:Nd$^{3+}$ laser ($\lambda=1.06$ $\mu$m) with energy $E=110$ mJ in the Q-switched regime ($\tau$ = 15 ns) and a pulse repetition frequency of 25 Hz, which was focused on the Ad and Pt targets. The substrate (HOPG) was mounted at a distance of 5 cm from the laser in the direction of  laser ablation plasma plume.

The chemical composition of the HOPG surface and Pd and Pt nanoparticles was controlled at the level of 0.1\% by means of X-ray photoelectron spectroscopy in the UHV analysis chamber ($p\approx5\times 10^{-11}$ Torr) of the Multiprobe~MXPS~RM~VТ~AFM-25 surface analysis system (Omicron). The size, shape and tunneling $IV$-curves of the nanoclusters were determined by analysing the Omicron UHV AFM/STM LF1 scanning tunneling microscope (STM) images. For this purpose, after the  attestation of the chemical composition the sample was \textit{ex situ} moved into the  probe microscopy chamber of UHV surface analysis system. STM images and $IV$-curves were obtained with the use of the same platinum tip at the values of feedback current of $I_0 = 1$~nA and bias voltage of $V_0 = 0.1$~V.

\section{Results}
\begin{figure*}[htb]%
a) {\includegraphics*[width=.35\textwidth,height=5.0cm]{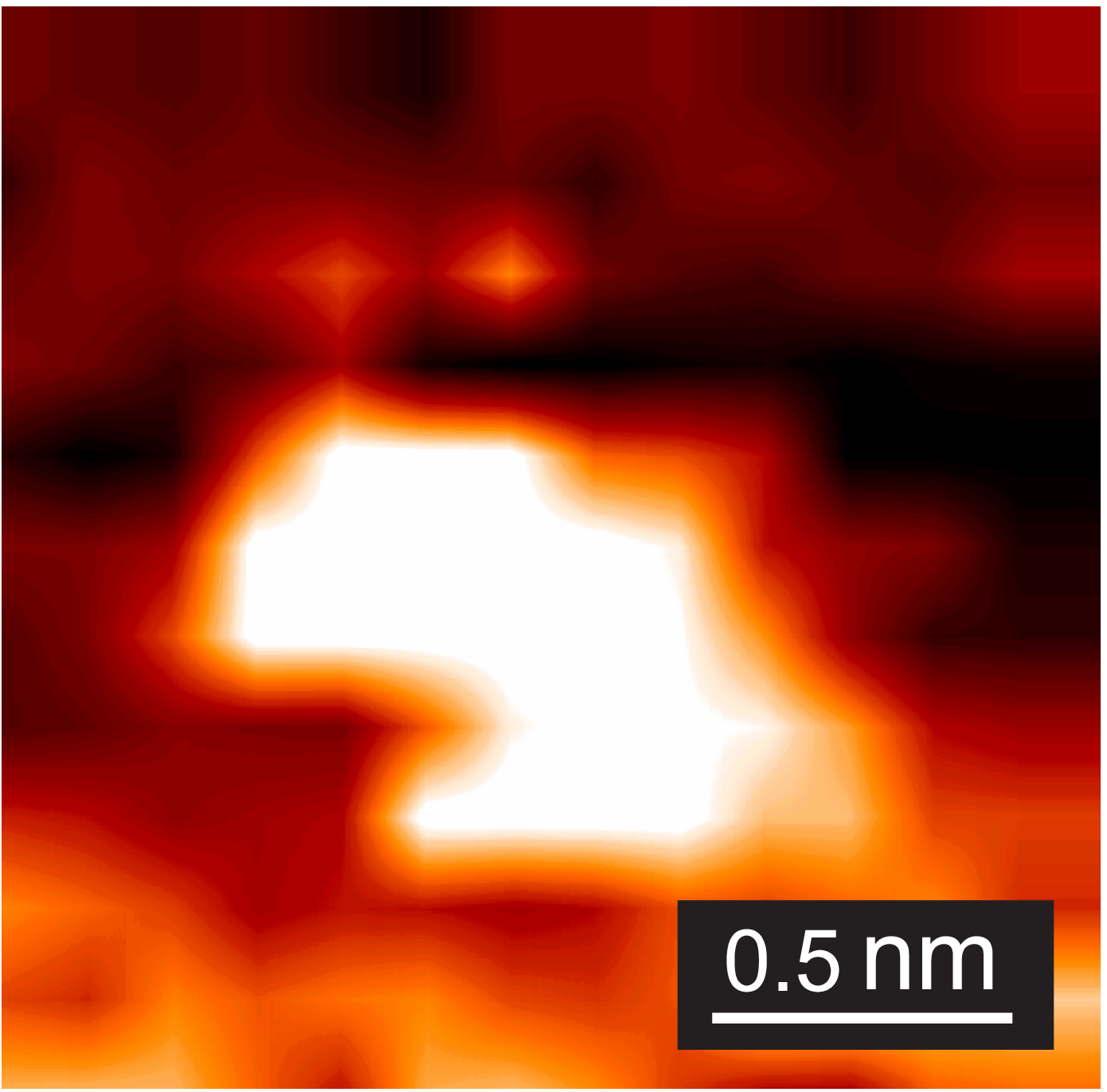}}\hfill{\includegraphics*[width=.47\textwidth,height=5cm]{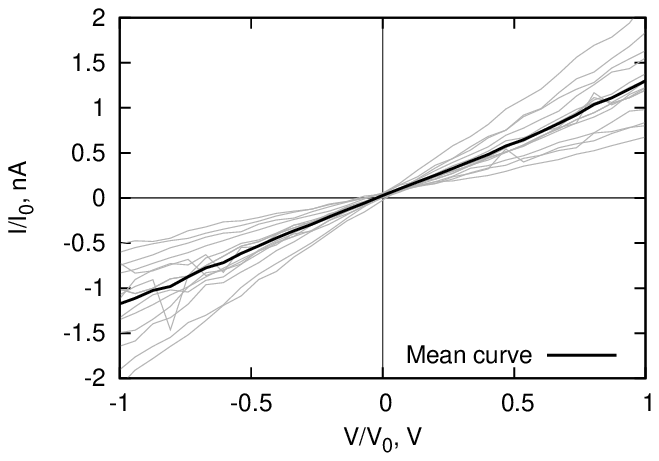}}\\[0.1cm]
b) {\includegraphics*[width=.35\textwidth,height=5.0cm]{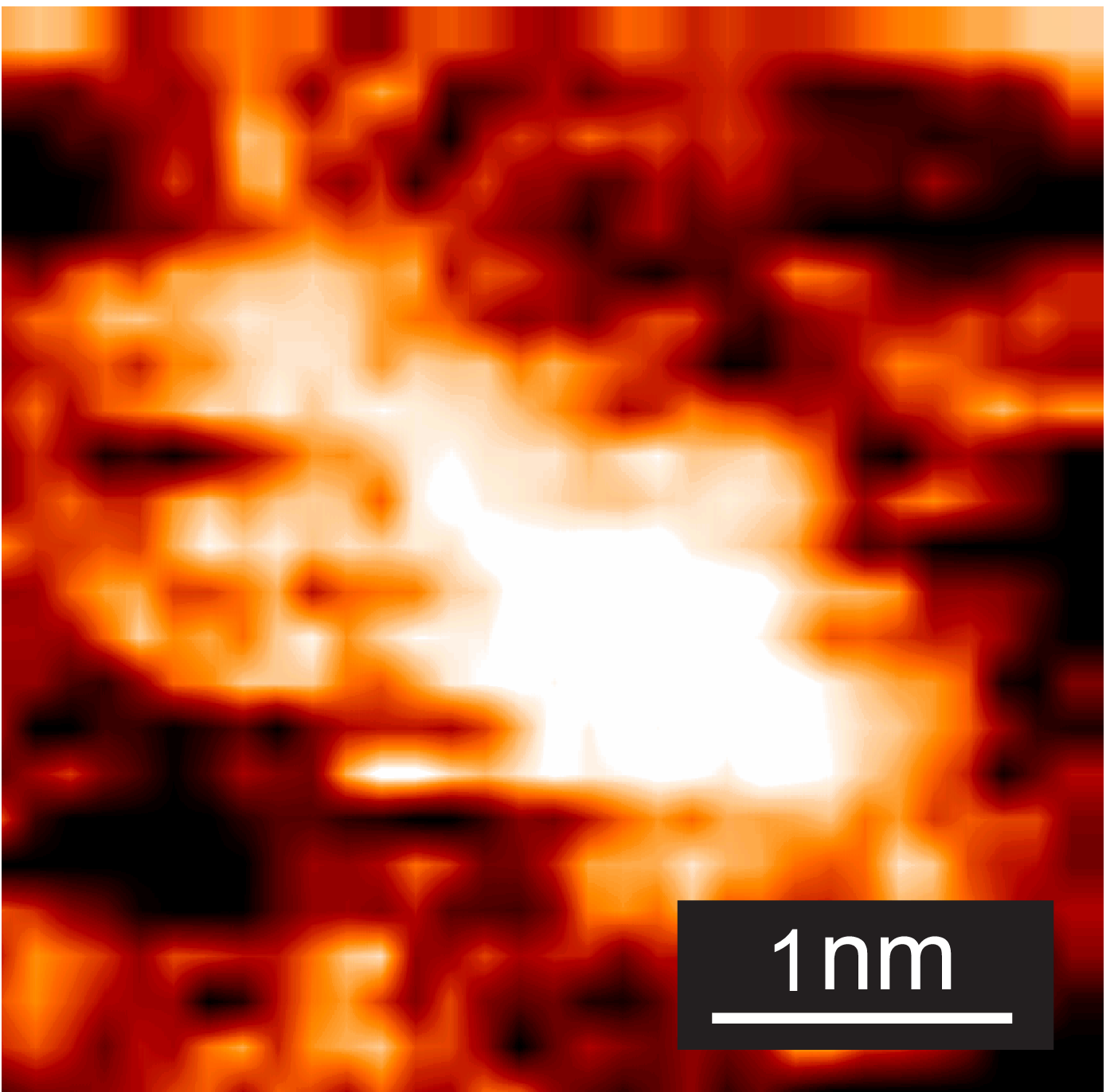}}\hfill{	\includegraphics*[width=.47\textwidth,height=5cm]{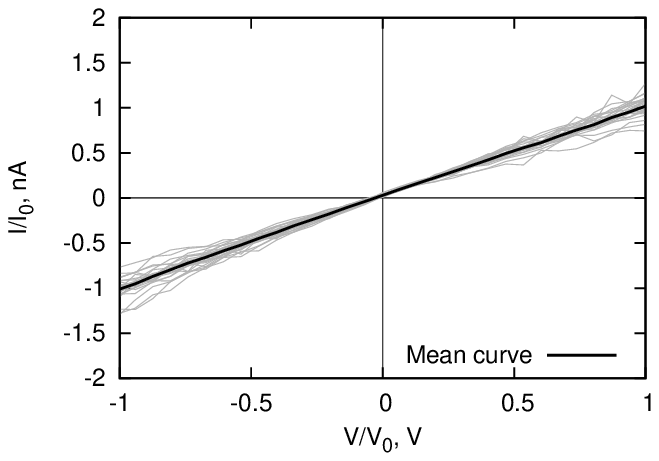}}\\
\caption{Typical STM images of PLD Pd(a) and Pt(b) nanoclusters with set of corresponded $IV$-curves measured in different points of each nanocluster. Corresponded volumes and values of averaged differential tunnel conductivities $\langle D\rangle =\langle\left.\frac{V_0}{I_0}\frac{d I}{d V}\right|_{V=0}\rangle$: a)~$v=0.28$~nm$^3$, $\langle D \rangle=1.45\pm0.37$;  b)~$v=3.04$~nm$^3$, $\langle D\rangle=1.01\pm0.07$}
\label{fig:1}
\end{figure*}

\begin{figure}[t]%
\begin{minipage}{0.5\linewidth}
\includegraphics*[width=0.90\linewidth]{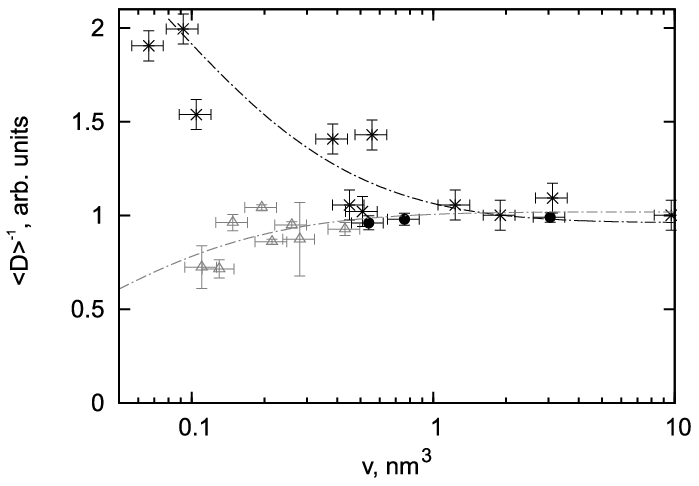}
\end{minipage}
\begin{minipage}{0.5\linewidth}
\includegraphics*[width=0.99\linewidth]{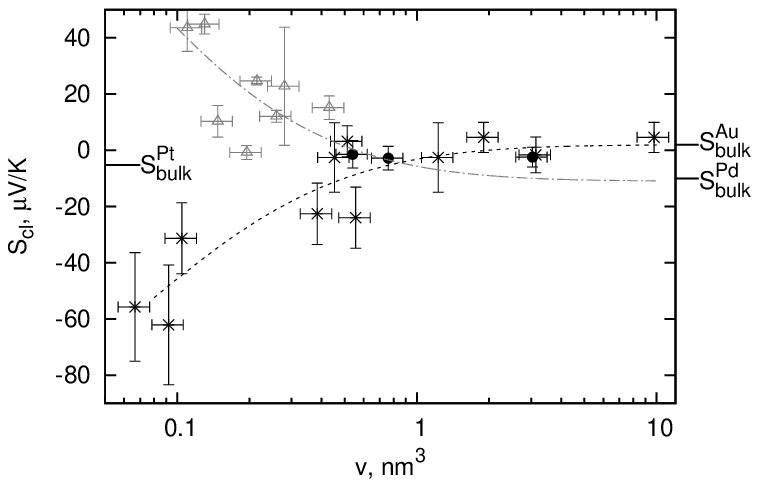}
\end{minipage}
\begin{minipage}{0.5\linewidth}
\centering a)
\end{minipage}
\begin{minipage}{0.5\linewidth}
\centering b)
\end{minipage}
	\caption{%
		a) The dependence of the mean differential tunneling conductance $\langle D\rangle =\langle\left.\frac{V_0}{I_0}\frac{d I}{d V}\right|_{V=0}\rangle$ and b) the dependence of thermoelectric power value $S_{cl}$ of PLD Au($\times$), Pd($\triangle$) and Pt($\bullet$) nanoclusters on their volume $v$.}
	\label{fig:2}
\end{figure}

Sets of tunneling $IV$-curves measured for typical PLD Pd and Pt nanoclusters are shown in Fig.~\ref{fig:1} (a-b). 
It can be seen that the tunneling $IV$-curves measured between the tip and Pd and Pt  nanoclusters are linear within the range $V/V_0=-0.1\div0.1$. As a result of analysis of series of differential tunneling $IV$-curves of each Pd and Pt nanocluster the dependences of mean differential tunneling conductance $\langle\left.\frac{dI}{dV}\right|_{V=0}\rangle$ on nanocluster's volume were recovered (see Fig.~\ref{fig:2} a)). For comparison the values  $\langle\left.\frac{dI}{dV}\right|_{V=0}\rangle$ for Au nanoclusters obtained previously \cite{Borisyuk2014} are also shown in Fig.~\ref{fig:2} a).
It can be seen that differential tunneling conductance in case of Pd nanoclusters is sensitive to their size as it was previously shown for Au nanoclusters \cite{Borisyuk2014} but shows qualitative difference in size dependence. The value of differential tunneling conductance for Pd nanoclusters increases with increasing of nanocluster size and reaches the value of $\langle\left.\frac{dI}{dV}\right|_{V=0}\rangle \frac{V_0}{I_0}\simeq1.4$ for Pd clusters of 0.1 nm$^3$ in volume whereas differential tunneling conductance for Au nanoclusters decreases with nanocluster's volume. Differential $IV$-curves of Pt nanoclusters in range of volumes 0.5$\div$3 nm$^3$ apparently doesn't shows any size dependence on size.  The choose of the volume for characterizing the size of nanoclusters was determined by the strongly irregular structure of nanoclusters formed by PLD on the substrate surface \cite{Borman2001} that lowers the accuracy of linear size determination. In this case the volume as the parameter that characterizes the size of nanoclusters allows the correct comparison of tunnel $IV$-curves of PLD nanoclusters of different materials. Before starting the discussion of the results obtained in terms of the thermoelectric properties of single nanoclusters let us discuss the essence of the proposed method of thermoelectric power value recovery of metal nanoclusters deposited on the substrate surface.

The technique of thermoelectric power value recovery is described in detail previously \cite{Borisyuk2014,Troyan2014}. As differential tunneling conductance of metallic samples measured by STS technique depends on density of the electronic states (DOS) and therefore contains information on the qualitative behaviour of DOS of the sample then for the ideal Fermi gas in low temperature approximation ($kT\ll E_F$) and low voltages $eV\ll E_f$ the averaged differential tunneling conductance $\langle\left.\frac{dI}{dV}\right|_{V=0}\rangle \frac{V_0}{I_0}$ of sample (nanocluster) can be connected with it's thermoelectric power value \cite{Borisyuk2014,Troyan2014} (which itself is connected with DOS of the sample \cite{ziman1979Eng}):
\begin{equation}
	\label{eq:1}
	\left(\left\langle\left.\frac{dI}{dV}\right|_{V=0}\right\rangle \frac{V_0}{I_0}\right)^{-1}\simeq 1-\frac{3e^2V_0}{1\pi^2k^2T}(S_{cl}(v)+S_t)
\end{equation}
where $I_0$ and $V_0$ are feedback current and voltage at a constant interaction mode, respectively, $T$ is temperature, $S_{cl}$ and $S_{t}$ are the local thermoelectric power of the investigated nanocluster, and STM tip respectively. Thus equation (\ref{eq:1}) allows us to restore the thermoelectric power value of the sample nanocluster using tunnel $IV$-curves measured by STM. Then, from the analysis of experimentally measured tunneling current-voltage characteristics of the test metal samples, with known thermoelectric power value of platinum probe $S_t = -5.28$ $\mu$V/K, one can restore the local value of the thermoelectric power value of the sample. The approximations used in this technique were discussed in original papers \cite{Borisyuk2014,Troyan2014}.

In this case in accordance with above-mentioned technique the dependence of averaged over the surface of the cluster differential tunneling conductance allows us to recover  thermoelectric power value of the cluster $S_{cl}$ of definite volume. The obtained dependencies of thermoelectic power value on cluster's size are shown in Fig.~\ref{fig:2} b).

It can be seen that thermoelectric power value of Pd nanoclusters increases with nanocluster's volume whereas thermoelectric power value of gold nanoclusters show qualitative difference and decreases with decreasing of cluster's volume. For Au and Pd nanoclusters of volume more than 10~nm$^3$ the thermoelectric power value is typical for bulk materials ($S_{Au}=1.94$~$\mu$V/K, $S_{Pd}=-5.28$~$\mu$V/K at room temperature\cite{Bass1985}). For nanoclusters of volume of about  0.1~nm$^3$ the absolute values of thermoelectric power of Au and Pd are several times greater in the absolute value compared with bulk materials ($S_{cl}^{Pd}\sim+50$~$\mu$V/K, $S_{cl}^{Au}\sim-70$~$\mu$V/K). Thermoelectric power value of Pt nanoclusters in volume ranges 0.5$\div$3 nm$^3$ remains close to bulk value.

The difference in differential tunneling conductance $\langle\left.\frac{dI}{dV}\right|_{V=0}\rangle \frac{V_0}{I_0}$ of Au and Pd nanoclusters might be caused by different electron structure of studied nanoclusters. Indeed, according to the paper \cite{Smith1974}, DOS behaviour at Fermi-level is different for these two metals that leads to qualitative difference in differential tunneling conductance behaviour of Au and Pd nanoclusters and therefore difference in thermoelectric power. In this case for differential tunneling conductance (and therefore thermoelectric power) of Pt nanoclusters the behaviour similar to Pd nanoclusters should be expected for clusters size less than 0.5 nm$^3$. Verification of this assumption will be the task of future research.

Thus, in this paper study of  thermoelectric power values of Pd and Pt nanoclusters pulse laser deposited onto HOPG surface and comparison with previous results on Au clusters is presented. recovering of the thermoelectric power value the of metal nanoclusters deposited onto the substrate. It was shown that thermoelectric power value and differential $IV$-curves $\langle\left.\frac{dI}{dV}\right|_{V=0}\rangle \frac{V_0}{I_0}$ of studied nanoclusters of volume more than 10~nm$^3$ is typical for bulk materials. At smaller volumes Au and Pd nanoclusters reveal qualitative difference in thermoelectric power dependence behaviour. 
Thermoelectric power of pulse laser deposited Pd nanoclusters increases with their volume and reaches $S_{cl}^{Pd}\sim+50$~$\mu$V/K at volume of 0.1 nm$^3$ whereas thermoelectric power of pulse laser deposited Au nanoclusters decreases and reaches $S_{cl}^{Pd}\sim-70$~$\mu$V/K at the same volume of nanocluster. Thermoelectric power value of Pt nanoclusters in volume ranges 0.5$\div$3 nm$^3$ remains close to bulk value.

\section*{Acknowledgements}
We are grateful to D.O. Filatov and D.A. Antonov for assistance in conducting the experiments. This work was supported by Russian Foundation for Basic Research (projects No.14-08-00487 a and No. 15-08-06153) and Ministry of Education of Russia (project No. 3.1803.2014/K).

\end{document}